\newif\ifdraft
\draftfalse
\documentclass[fleqn,10pt]{wlscirep}
\usepackage[utf8]{inputenc}
\usepackage[T1]{fontenc}
\usepackage{multirow}
\ifdraft
\newcommand{\added}[1]{\textcolor{blue}{#1}}
\newcommand{\deleted}[1]{\textcolor{red}{\stkout{#1}}}

\newcommand{\deletedfloat}[1]{}
\newcommand{\commented}[1]{\textcolor{blue}{#1}}
\else
\newcommand{\added}[1]{#1}
\newcommand{\deleted}[1]{}

\newcommand{\deletedfloat}[1]{}
\newcommand{\commented}[1]{}
\fi

\title{Towards Objective Gastrointestinal Auscultation: Automated Segmentation and Annotation of Bowel Sound Patterns}

\author[1,3]{Zahra Mansour}
\author[2]{Verena Uslar}
\author[2]{Dirk Weyhe}
\author[3]{Danilo Hollosi}
\author[1,*]{Nils Strodthoff}
\affil[1]{Division AI4Health, Department for Health Services Research, Faculty of Medicine and Health Sciences, Carl von Ossietzky Universität Oldenburg, 26129 Oldenburg, Germany}
\affil[2]{University Clinic for Visceral Surgery, Faculty of Medicine and Health Sciences, Carl von Ossietzky Universität Oldenburg, 26121 Oldenburg, Germany}
\affil[3]{Fraunhofer IDMT, Institute Part HSA, 26129 Oldenburg, Germany}
\affil[*]{nils.strodthoff@uol.de}


\begin{abstract}
Bowel sounds (BS) are typically momentary and have low amplitude, making them difficult to detect accurately through manual auscultation. This leads to significant variability in clinical assessment. Digital acoustic sensors allow the acquisition of high-quality BS and enable automated signal analysis, offering the potential to provide clinicians with both objective and quantitative feedback on bowel activity. This study presents an automated pipeline for bowel sound segmentation and classification using a wearable acoustic SonicGuard sensor.\\
BS signals from 83 subjects were recorded using a SonicGuard sensor. Data from 40 subjects were manually annotated by clinical experts and used to train an automatic annotation algorithm, while the remaining subjects were used for further model evaluation. An energy-based event detection algorithm was developed to detect BS events. Detected sound segments were then classified into BS patterns using a pretrained Audio Spectrogram Transformer (AST) model. Model performance was evaluated separately for healthy individuals and patients.\\ 
The best configuration used two specialized models, one trained on healthy subjects and one on patients, achieving (accuracy: 0.97, AUROC: 0.98) for healthy group and (accuracy: 0.96, AUROC: 0.98) for patient group. The auto-annotation method reduced manual labeling time by approximately 70\%, and expert review showed that less than 12\% of automatically detected segments required correction.\\ 
The proposed automated segmentation and classification system enables quantitative assessment of bowel activity, providing clinicians with an objective diagnostic tool that may improve the diagnostic of gastrointestinal function and support the annotation of large-scale datasets.
\end{abstract}

\begin{document}

\flushbottom
\maketitle
\section*{Introduction}
\begin{figure}[t]
\centering
\includegraphics[width=0.9\textwidth]{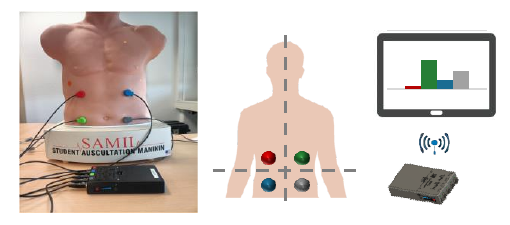}
\caption{SonicGuard wearable acoustic sensing platform. Four sensors are placed on the abdominal quadrants (Right Upper Quadrant (RUQ), Left Upper Quadrant (LUQ), Right Lower Quadrant (RLQ), and Left Lower Quadrant (LLQ)) to continuously capture bowel sounds. The collected signals are transmitted  to the processing platform, which forwards them to the companion mobile application for automated analysis and real-time user feedback.}
\label{fig1}
\end{figure}

\begin{figure}[t]
\centering
\includegraphics[width=0.9\textwidth]{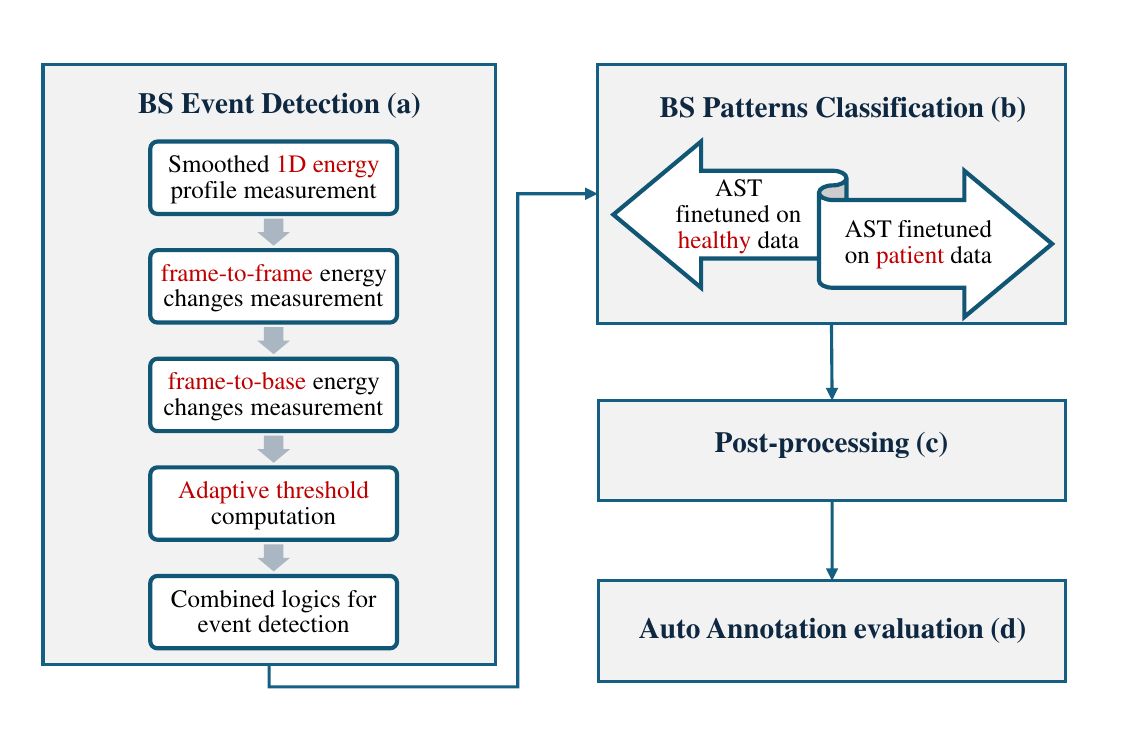}
\caption{The methodology used in the auto-annotation algorithm starting from BS event detection by measuring the changes in the energy, followed by BS pattern classification using AST model, then adjusting the segment and finish by comparing the manual created label to the auto labels.}
\label{fig2}
\end{figure}

Bowel sounds (BS) are sporadic acoustic events produced by the contraction of the smooth muscles in the abdominal wall and the movement of gas and fluids through the intestines during digestion. Thus, BS reflect gastrointestinal motility and overall digestive function \cite{Breum2015}. Abdominal auscultation was first proposed as a clinical technique for assessing gastrointestinal activity in the early 20th century \cite{Cannon1905}, and since then, clinicians have sought to understand the relationship between BS, gastrointestinal physiology, and disease states \cite{Talley2013}. In abdominal examination practice, physicians typically place a stethoscope over the four abdominal quadrants and listen for several minutes to determine whether BS are present or absent, which is interpreted as an indication of bowel obstruction or postoperative ileus. However, this assessment is subjective and lacks quantitative measurements\cite{Breum2015, Koenig2011}.\\
Traditional auscultation of BS is challenged by the inherently intermittent nature of BS signals. In contrast to heart or lung sounds, which present rhythmic or continuous patterns, BS occur at unpredictable intervals and thus may be missed during short listening sessions \cite{Baronetto2024}. Consequently, clinicians often need to listen for several minutes across multiple abdominal quadrants to detect acoustic events, increasing clinical workload and reducing practicality in busy care settings \cite{Koenig2011}. Moreover, bowel sound events are characterized by very short durations (from milliseconds to a few seconds) and low energy, which makes them difficult to detect reliably with the human ear and renders auscultation highly subjective \cite{Yu2024}. These properties reduce sensitivity and inter-examiner consistency, and underscore the need for automated, quantitative methods that can provide objective and reproducible assessment beyond listening.\\
The emergence of high-fidelity digital stethoscopes and wearable acoustic sensors now enables continuous recording of bowel sounds with sufficient quality for computational analysis \cite{choudry2022} \cite{Mansour2024}. This overcomes the limitations of traditional auscultation and allows researchers to apply automated methods for detecting and characterizing bowel sound activity. However, progress in this field remains constrained by the scarcity of large, labeled datasets, making bowel sound analysis more challenging than other physiological acoustic signals.
Research on automated BS analysis has evolved from basic signal processing toward advanced machine learning and deep learning methods. Early studies focused mainly on detection of bowel sound events, using filtering or time–frequency representations to isolate BS from noise, such as wavelet-based stationary filtering \cite{hadjileontiadis1999}, STFT-based analysis \cite{ficek2021}, and fractal or nonlinear feature extraction \cite{dimoulas2007}. With the rise of machine learning, classical models such as SVMs and neural networks were used to detect BS segments and distinguish them from artifacts \cite{yin2018}. More recent work shifted toward deep learning: CNN-based detectors can now identify BS events in continuous recordings and have been applied to behavioral or digestion-state monitoring \cite{wang2022}. A newer research direction is classification of bowel sound patterns, aiming not only to detect BS events but also to categorize them into physiologically meaningful types. Early attempts included wavelet neural networks \cite{dimoulas2007} and spectrogram-based clustering \cite{du2018}. Recent advances demonstrate the use of pretrained audio models—such as Wav2Vec and CNN-based architectures—to improve recognition accuracy and reduce annotation requirements \cite{Yu2024,Baronetto2024Multiscale,Mansour2026}. Despite these advances, only a limited number of studies classify BS into multiple clinically relevant patterns \cite{Mansour2026}.
\added{As highlighted in the studies reviewed above, existing researches have primarily focused on either detecting bowel sound segments or classifying predefined segments into bowel sound patterns, see, e.g., \cite{Mansour2026}. However, no research has yet automated the complete pipeline, from bowel sound auscultation to quantitative pattern analysis, within a single unified framework. This lack of an end-to-end system that integrates automated bowel sound detection with pattern-level classification represents a clear research gap.}\\
In this study, we build upon a previously developed wearable acoustic sensor \cite{Mansour2024} and apply it to record bowel sounds (BS) from more than 80 subjects, including both healthy individuals and patients. Using these recordings, we developed an automated analysis framework capable of detecting bowel sound activity and categorizing events into four main BS patterns. The algorithm was evaluated using expert-reviewed annotations, and its performance was benchmarked against manually labeled data. The findings highlight the potential of such automated analysis to provide clinicians with quantitative and objective feedback regarding gastrointestinal activity, supporting more informed decision-making in everyday clinical practice.

\begin{figure}[t]
\centering
\includegraphics[width=0.7\textwidth]{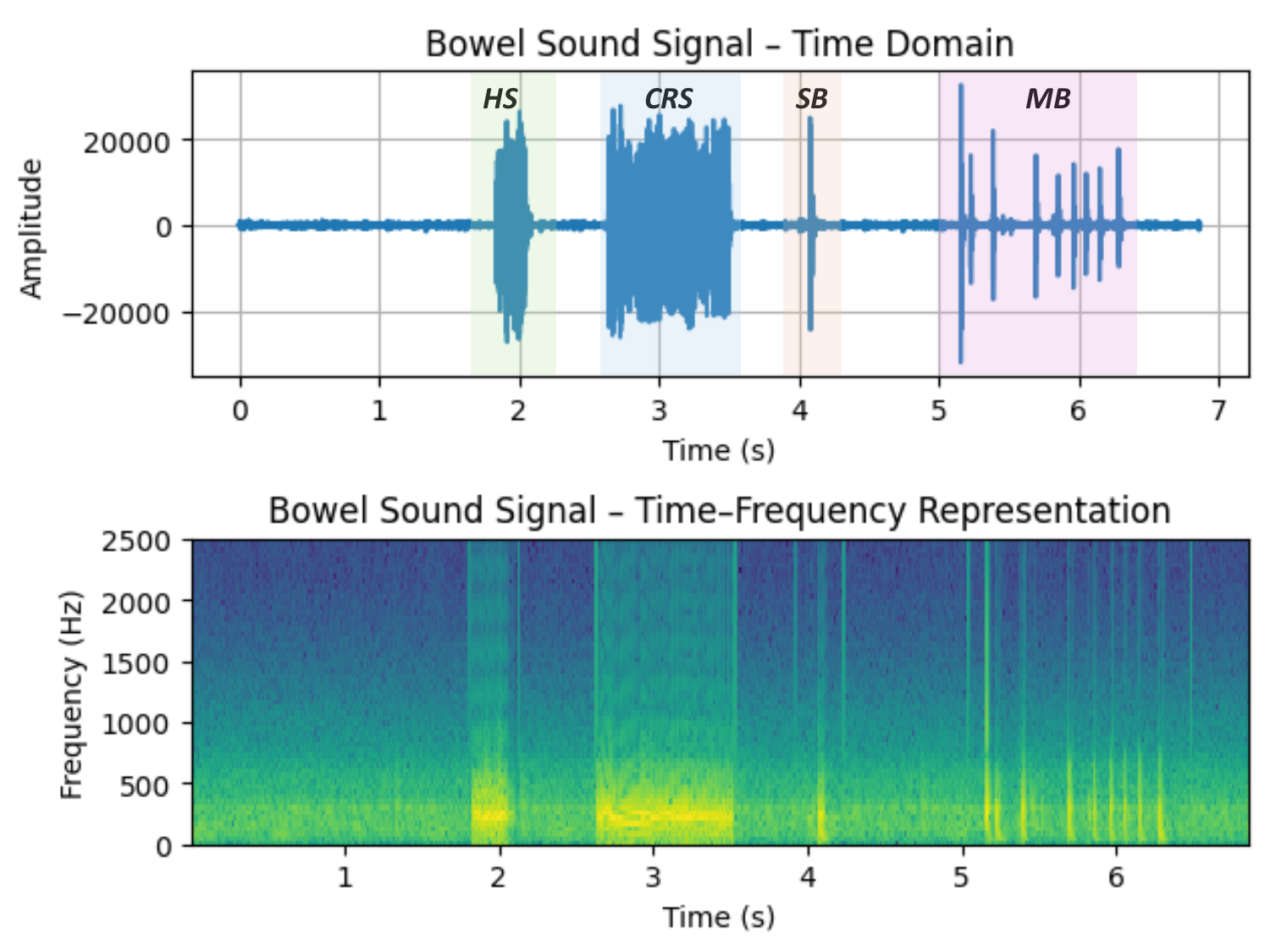}
\caption{The four BS patterns investigated in this study in time and frequency domain, starting from Harmonic Sound (HS), followed by Continuous Random Sound (CRS), Single Burst (SB) and Multiple Burst (MB)}
\label{fig3}
\end{figure}
\section*{Materials and methods}
\subsection*{Dataset}
\paragraph{Recording protocol} The research protocol underlying the data recordings was approved by the Medical Ethics Committee of the University of Oldenburg (2022-056). Before participation, subjects were informed about the purpose of the study and consented to the use of the collected data in an anonymized form. 
During the recording session, each subject was asked to lie down in a supine position. the BS were then recorded from the four abdominal quadrants (Right Upper Quadrant (RUQ), Left Upper Quadrant (LUQ), Right Lower Quadrant (RLQ), Left Lower Quadrant (LLQ)) as it shown in Fig \ref{fig1}, using SonicGuard wearable multichannel acoustic sensor\cite{Mansour2024} . Seven minutes of BS were recorded from each quadrant with a total of 28 minutes of recording for each subject. \\

\paragraph{Participant Information} 
A total of more than 40 hours of bowel sound recordings were acquired from 84 subjects, including 36 patients diagnosed with gastrointestinal (GI) diseases and 48 healthy controls, at PIUS Hospital in Oldenburg, Germany. Participants ranged in age from 20 to 83 years, comprising 40 females and 44 males, with body mass index (BMI) values ranging from 18 to 40~kg/m$^2$. The patient group included individuals with colon cancer, Crohn’s disease, and postoperative gastrointestinal disorders.

\paragraph{Data Annotation and Bowel Sound Patterns}

The BS patterns analyzed in this study were defined in detail in previous work~\cite{Mansour2026}. They align with classifications commonly used in abdominal auscultation and prior literature, covering all sound events observed in the dataset. The patterns, examples for each of them are shown in Fig.~\ref{fig3}, are summarized as follows:\\
\noindent\textbf{Single Burst (SB)} Short, isolated impulses (10–30~ms) likely caused by small intestinal contractions or splashes of the fluids inside the luminal content.\\
\noindent\textbf{Multiple Burst (MB)} Groups of SBs separated by short, irregular gaps (40–1500~ms total duration), possibly produced by fluids movements within the intestine.\\
\noindent\textbf{Continuous Random Sound (CRS)} Continuous, clustered rumbling sounds (200–4000~ms) with no silent intervals, caused by the movement of gas-filled luminal contents or intestinal contents passing through sections of the intestine with variable diameters.\\
\noindent\textbf{Harmonic Sound (HS)} The least frequent pattern, featuring three to four harmonic frequency components (50–1500~ms), associated with stenosis.

\noindent The recordings from 40 subjects (18 patients and 22 healthy controls) were manually annotated into the four BS patterns described above by a member trained under the supervision of medical staff using the Audacity software, based on both the waveform and spectrogram characteristics as well as auditory evaluation. The remaining 43 recordings were used for the evaluation of the auto-annotation algorithm.

\begin{figure}[t!]
\centering
\includegraphics[width=0.9\textwidth]{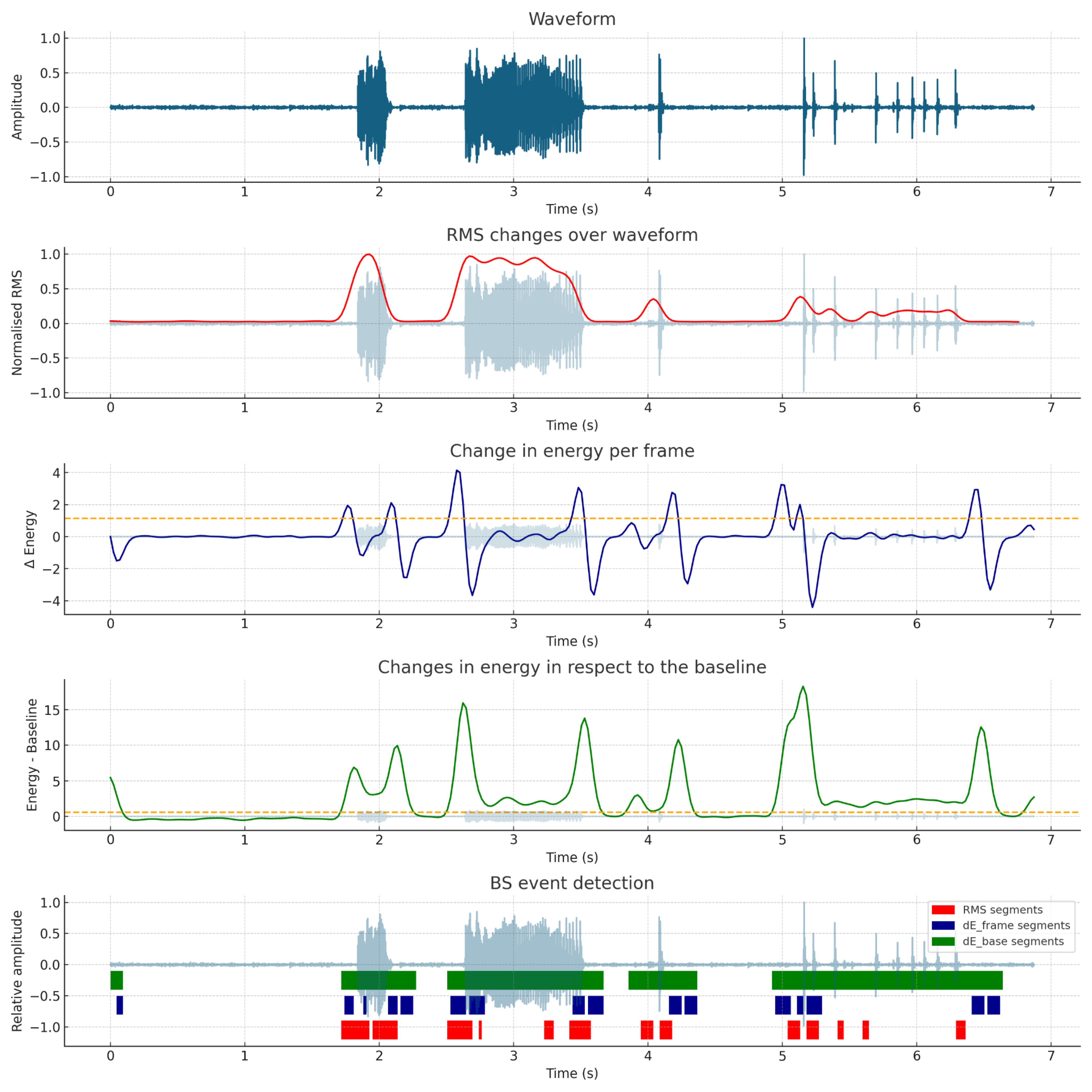}
\caption{Illustration of the proposed bowel sound (BS) event detection features. The first row shows the bowel sound waveform in the time domain. The second row presents the temporal variations of the RMS amplitude along the waveform with the corresponding adaptive threshold. The third row depicts the changes in signal energy within each short frame together with its adaptive threshold. The fourth row shows the changes in frame energy relative to the baseline energy of the signal and the associated adaptive threshold. The final row displays the detected BS events using different features: RMS-based detection (red), energy variation within frames (blue), and energy variation relative to the baseline energy (green).}
\label{fig4}
\end{figure}

\subsection*{Methodology}
The methodology used in the auto-annotation algorithm is visualized in Fig \ref{fig2}. Below, we provide a detailed description for each of its main components.
\subsubsection*{BS Event Detection}
Unlike quasi-periodic biomedical acoustic signals such as heart sounds, bowel sounds (BS) exhibit substantial morphological variability and can be categorized into four distinct patterns, as described above. These patterns differ in both temporal and spectral characteristics, posing challenges for event detection using a single feature or criterion. Short-duration, impulsive BS events (SB and MB) are difficult to detect using conventional energy-based approaches that rely on changes relative to baseline energy, while longer-duration, cluster-like patterns (CRS and HS) may not be fully captured by methods based solely on frame-level energy or RMS amplitude.
\noindent To address these challenges, A customized bowel sound (BS) event detection algorithm was developed using short-time analysis with non-overlapping frames of 1\,ms duration.

\noindent Let $x_i[n]$ denote the discrete-time signal samples within the $i$-th frame, where $n = 1, \ldots, N$, where $N$ denotes the number of tokens in a frame. For each frame, time-domain features characterizing intra-frame amplitude and energy variations are extracted to identify candidate event onset and offset boundaries.

\noindent The frame-level root mean square (RMS) amplitude and signal energy are defined as
\begin{equation}
\mathrm{RMS}_i = \sqrt{\frac{1}{N}\sum_{n=1}^{N} x_i^2[n]},
\end{equation}
\begin{equation}
E_i = \sum_{n=1}^{N} x_i^2[n].
\end{equation}

\noindent To reduce inter-subject variability and improve robustness to background noise, \added{the energy representation is first converted to the decibel scale using a global reference equal to the maximum spectral magnitude within the recording, ensuring consistent normalization across the signal. A frame-level energy profile is then obtained by averaging across frequency bins and smoothing over time. The baseline is defined as the median value of the smoothed energy curve, which represents the typical non-event activity level used for normalization. The normalized RMS and normalized energy are computed as}
\begin{equation}
\mathrm{RMS}_i^{\mathrm{norm}} = \frac{\mathrm{RMS}_i}{\overline{\mathrm{RMS}}_{\mathrm{baseline}}},
\end{equation}
\begin{equation}
E_i^{\mathrm{norm}} = \frac{E_i}{\overline{E}_{\mathrm{baseline}}},
\end{equation}
where $\overline{\mathrm{RMS}}_{\mathrm{baseline}}$ and $\overline{E}_{\mathrm{baseline}}$ denote the mean RMS and energy values calculated over the baseline segment, respectively.

In parallel, temporal variations in normalized frame-level energy are analyzed across consecutive frames to characterize sustained acoustic activity.
\added{BS detection is performed using a joint decision strategy based on RMS amplitude and energy variation criteria. Specifically, the onset of a new segment is defined when both the normalized RMS value and the frame-to-frame energy difference exceed a predefined threshold set to the median of their respective frame-wise distributions. To ensure temporal continuity and prevent erroneous fragmentation (particularly in cluster-like BS patterns where intra-frame energy fluctuations may be misleading), the segment is maintained as long as the energy relative to the baseline remains above the same median-based threshold defined among the entire record. The end of a segment is declared only when all three parameters (RMS, frame-to-frame energy difference, and energy relative to the baseline) fall below the threshold.}
The joint use of intra frame amplitude dynamics and inter-frame energy continuity enables reliable detection of BS events with heterogeneous temporal patterns within a unified detection framework.

\subsubsection*{BS pattern classification}
After detecting signal segments containing BS, a dedicated classification stage was applied to determine the BS type among the four predefined patterns. Recent findings \cite{Mansour2026} have demonstrated that pretrained deep learning models provide superior performance for BS pattern classification by leveraging transferable acoustic representations learned from large-scale audio corpora. Based on this evidence, two state-of-the-art pretrained models were evaluated in this study: the Wav2Vec model and the Audio Spectrogram Transformer (AST). \\
\noindent An Audio Spectrogram Transformer (AST) classifier \cite{Gong2021} operates on two-dimensional log Mel-spectrograms by partitioning each spectrogram into non-overlapping patches, which are linearly embedded and processed using a transformer architecture with self-attention mechanisms, which enables each time–frequency patch to attend to all other patches across the spectrogram, with positional embeddings preserving temporal order. The network was initialized with weights pretrained on the AudioSet dataset \cite{Gemmeke2017}, allowing the extraction of generic and robust audio representations prior to task-specific finetuning.\\
\noindent In addition, a Wav2Vec-based classifier was evaluated for BS pattern classification \cite{Baevski2020}. Unlike AST, Wav2Vec operates directly on raw audio waveforms and learns hierarchical latent speech and audio representations through self-supervised pretraining. The model captures fine-grained temporal information and long-context dependencies without relying on explicit time–frequency transformations, making it well suited for impulsive and non-stationary acoustic events such as bowel sounds. The pretrained Wav2Vec encoder was finetuned end-to-end using the detected BS segments to adapt its learned representations to the BS pattern classification task.
\begin{figure}[t!]
\centering
\includegraphics[width=0.9\textwidth]{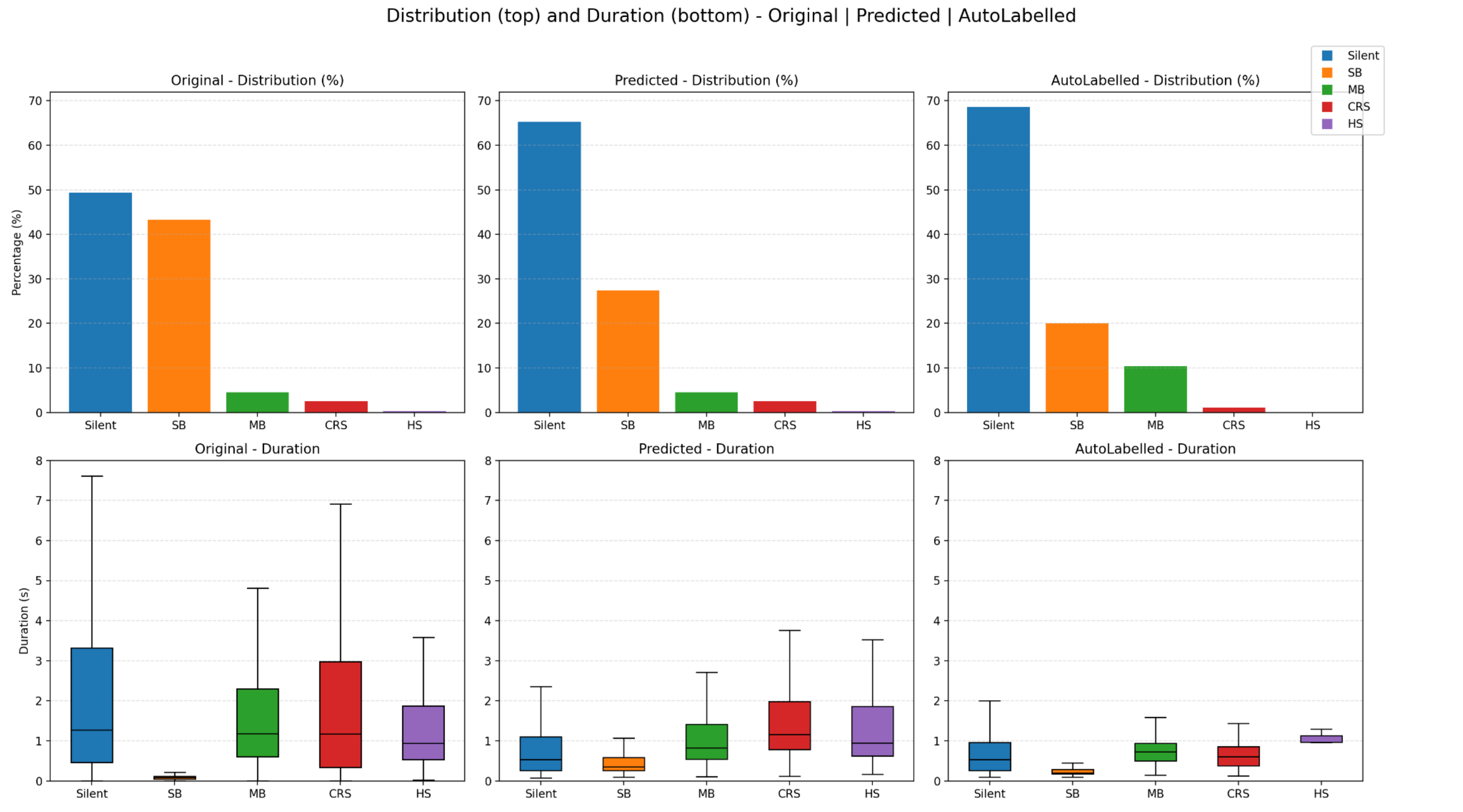}
\caption{The differences in distribution (first row) and duration (second row) between the bowel sound patterns in the data that has been manually labeled titled as Original, and the same data if it is labeled using the auto annotation technique titled predicted and the group of patient that we do not have the manual label for and we have created the labels using the auto annotation algorithm labeled auto annotation, by using the model that has been trained on healthy data only and tested on healthy data too.}
\label{fig5}
\end{figure}
For each pretrained architecture, three cohort-specific models were finetuned: (i) Healthy-only, (ii) Patient-only, and (iii) a combined Healthy+Patient model. All models employed identical preprocessing (128-band log Mel-spectrograms with a 25~ms window and 10~ms hop size for AST; raw waveform input for Wav2Vec), per-band mean–variance normalization where applicable, the same training, validation, and test statified subject-level, split (70\%, 15\%, and 15\%), and matched hyperparameters. This controlled setup ensured that performance differences reflected cohort-related and domain-specific effects rather than training variability.

Model performance was quantified using the area under the receiver operating characteristic curve (AUROC), a threshold-independent metric of classification discrimination \cite{mcdermott2024}. Each model was evaluated on three test sets (Healthy-only, Patient-only, and Mixed) to assess within-domain performance, cross-domain generalization, and robustness to cohort shifts.

\subsubsection*{Post-processing}
After detecting the classification, the pipeline applies a temporal refinement stage. Detected events are ordered chronologically and time gaps between adjacent events are optionally filled with a default label (non-BS segment) when gaps exceed a minimum duration threshold (100 msec). A post-processing merging procedure is applied in which two or more temporally adjacent segments assigned the same predicted label are merged into a single segment. This step mitigates the artificial splitting of continuous events into smaller segments caused by frame-level fluctuations, resulting in longer, clinically interpretable sequences that represent consistent BS activity.

\subsubsection*{Evaluation}

The performance of the proposed automatic annotation algorithm was evaluated using three complementary data groups, designed to assess accuracy, usability, and statistical consistency.

\noindent First, manually annotated data from 12 unseen subjects were used as the reference standard to quantitatively evaluate the auto-annotation performance. The automatically generated labels were compared against the manual annotations in terms of event count, temporal distribution, and event duration, allowing direct assessment of agreement between the two labeling approaches.

\noindent Second, the remaining data from 35 subjects were used to assess the physiological plausibility and consistency of the auto-annotated labels in the absence of manual ground truth. Statistical features extracted from the automatically labeled bowel sound events—such as event duration, inter-event intervals, and temporal distribution—were compared with corresponding characteristics reported for normally annotated bowel sound data, to verify that the generated labels conformed to expected patterns.

\noindent Finally, data from 10 subjects were used to evaluate the practical utility of the auto-annotation system in a semi-automated labeling workflow. For this group, automatic labels were first generated and then reviewed by an expert annotator. The required annotation time and the number of manual adjustments applied to the auto-generated labels were recorded, providing a measure of annotation efficiency and reduction in expert workload.

\begin{table}[t]
\caption{Performance comparison between Wav2Vec~2.0 and AST models using Accuracy (ACC) and Area Under the Curve (AUROC).}
\label{tab:model_comparison}
\centering
\begin{tabular}{llcccc}
\toprule
\textbf{Train} & \textbf{Test} &
\multicolumn{2}{c}{\textbf{Wav2Vec~2.0}} &
\multicolumn{2}{c}{\textbf{AST}} \\
& & ACC & AUROC & ACC & AUROC \\
\midrule

\multirow{3}{*}{Healthy}
& \textbf{Healthy}  & 0.82 & 0.74 & \textbf{0.97} & \textbf{0.98} \\
& Patients          & 0.77 & 0.74 & 0.76 & 0.82 \\
& All               & 0.79 & 0.73 & 0.87 & 0.94 \\
\midrule

\multirow{3}{*}{Patients}
& Healthy           & 0.67 & 0.61 & 0.91 & 0.90 \\
& \textbf{Patients} & 0.75 & 0.69 & \textbf{0.96} & \textbf{0.98} \\
& All               & 0.73 & 0.66 & 0.93 & 0.95 \\
\midrule

\multirow{3}{*}{All}
& Healthy           & 0.73 & 0.80 & 0.94 & 0.93 \\
& Patients          & 0.75 & 0.81 & 0.94 & 0.96 \\
& \textbf{All}      & 0.77 & 0.83 & \textbf{0.94} & \textbf{0.98} \\
\bottomrule
\end{tabular}
\end{table}

\begin{figure}[t!]
\centering
\includegraphics[width=0.9\textwidth]{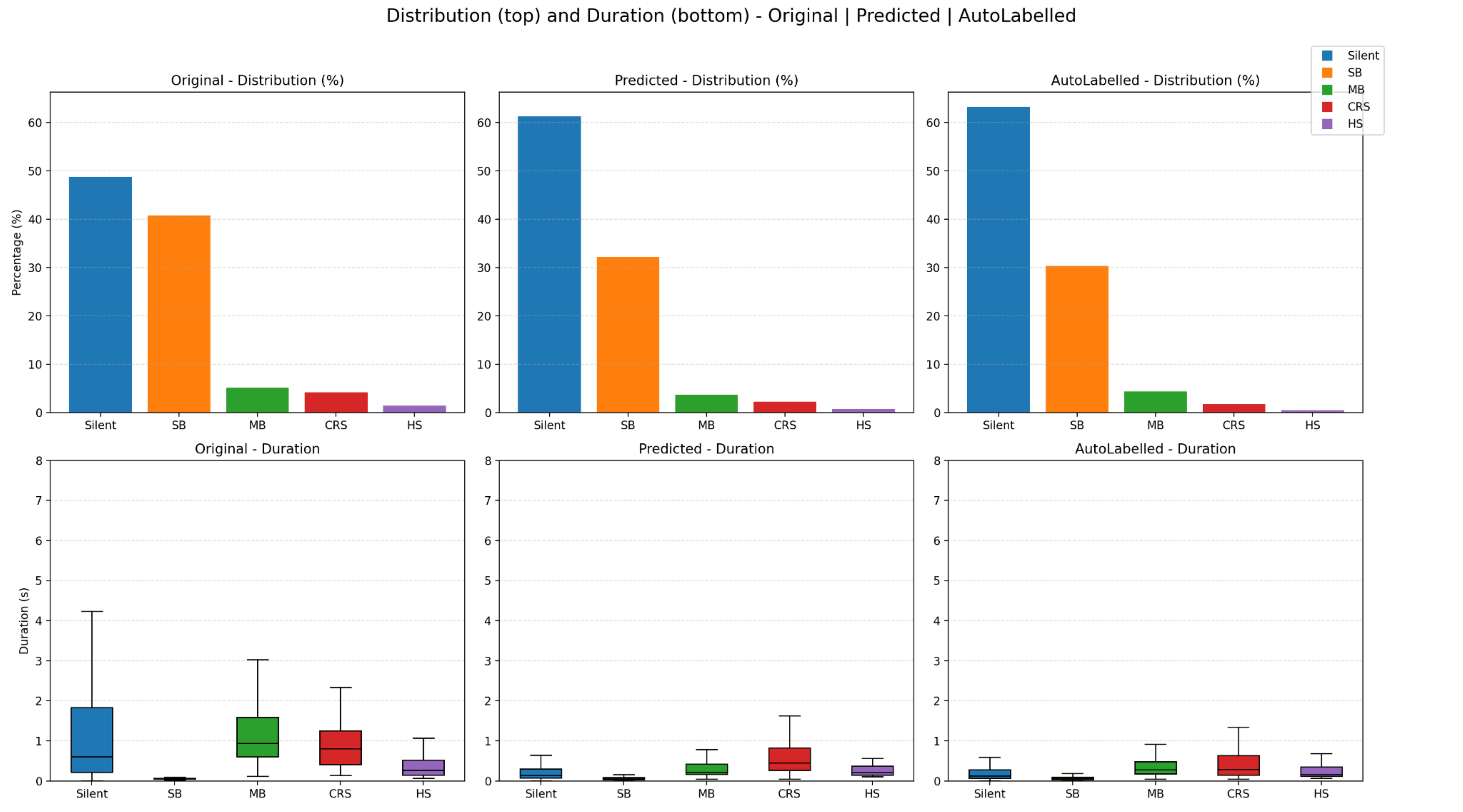}
\caption{The differences in normalized counts (first row) and duration (second row) between the bowel sound patterns between ground truth labels  (Original; left panel),  model predictions for the same set of samples (Predicted; center panel) and auto-annotated samples (AutoLabelled, right panel).}
\label{fig6}
\end{figure}

\section*{Results and Discussion}
\subsection*{BS event detection }
Since the BS event detection algorithm is based on a combination of multiple features, Fig.~\ref{fig4} illustrates how each feature captures different aspects of the various BS patterns. For example, the RMS-based feature is effective in detecting the onset and offset of BS events, particularly for peak-like patterns such as SB and MB. However, it struggles to detect the continuous portion of cluster-like events (MB and CRS), as these signals maintain a relatively constant amplitude over longer durations, leading RMS-based methods to incorrectly infer event termination. Similar difficulties are observed for algorithms relying on changes in energy within individual frames, which may identify event boundaries but fail to capture the full temporal extent of prolonged events. In contrast, incorporating the combination of features RMS-based detection, energy variation within frames, and energy variation relative to the baseline enables not only accurate detection of event onset and offset but also reliable identification of the entire BS event as a single continuous occurrence.
\subsection*{BS pattern classification}
Two pretrained models were evaluated for bowel sound (BS) pattern classification: Wav2Vec~2.0 and the Audio Spectrogram Transformer (AST). Each model was finetuned using three different training subsets (Healthy, Patients, and All). Classification performance in terms of accuracy (ACC) and area under the ROC curve (AUROC) is summarized in Table~\ref{tab:model_comparison}.

For both architectures, the highest performance within each training subgroup was achieved when the model was evaluated on test data drawn from the same cohort. Specifically, the model finetuned on healthy data achieved its best performance when tested on the healthy cohort using AST (ACC = 0.97, AUROC = 0.98). Similarly, the model finetuned on patient data performed best when evaluated on the patient test set, again using AST (ACC = 0.96, AUROC = 0.98). For the model finetuned on the combined dataset, the best results were obtained on the mixed test set with AST (ACC = 0.94, AUROC = 0.98).

These results demonstrate that morphological and structural differences between bowel sounds recorded from healthy subjects and patients directly influence classification performance. This cohort dependency motivated the adoption of cohort-specific classifiers within the automatic annotation framework, with model selection conditioned on whether the input data originate from healthy subjects or patients. On the other hand, the AST model trained on the combined cohort still shows a reasonable performance on both sub-cohorts.

Across all training and test subgroups, AST consistently outperformed Wav2Vec~2.0. Consequently, AST was selected as the default classifier in subsequent experiments.

\subsection*{BS auto-annotation framework}
We demonstrate possible application scenarios for the auto-annotation framework through two exemplary use case.
\subsubsection*{Use case 1: Automatic extraction of BS patterns characteristics}
To evaluate the performance of the proposed BS auto-annotation framework, we distinguish three subgroups. The \textit{Original} (org) group consists of manually annotated bowel sound labels from 40 subjects and serves as the reference standard. The \textit{Predicted} (pred) group includes labels automatically generated by the BS auto-annotation framework for the same manually annotated recordings, enabling direct comparison between manual and automated labeling. The \textit{Auto} group represents fully automatically generated labels obtained from recordings without manual annotations (43 subjects), and is used to assess the framework’s behavior on unlabeled data.
Results are presented separately for healthy subjects (Fig.~\ref{fig5}) and patient subjects (Fig.~\ref{fig6}) to highlight cohort-specific differences in bowel sound distributions and temporal characteristics.
For the healthy dataset, the original class distribution differs slightly from that of the patient cohort, with substantially fewer HS samples. This observation is expected, as HS patterns are more commonly associated with pathological conditions such as intestinal stenosis (Fig.~\ref{fig5}). The highest prevalence was consistently observed for the \textit{None} class (Original: 49\%, Predicted: 65\%, Auto: 68\%), followed by SB (Original: 43\%, Predicted: 28\%, Auto: 20\%), MB (Original: 5\%, Predicted: 5\%, Auto: 10\%), CRS (Original: 2.9\%, Predicted: 2.9\%, Auto: 1.99\%), and the rarest class, HS (Original: 0.1\%, Predicted: 0.1\%, Auto: 0.01\%). Similar to the patient dataset, the correct ordering of class prevalence was preserved by the automatic annotation framework.

In addition, BS patterns in healthy subjects originally tend to exhibit longer durations compared with patient data, particularly for MB, CRS, and HS events. This increase in event duration was successfully captured by the auto-annotation technique, despite its general tendency to slightly underestimate BS event lengths relative to manual labeling.
\begin{table}[t]
\caption{Class-wise comparison between automatically generated and expert-adjusted bowel sound labels (overall mean across 10 subjects and all quadrants).}
\label{tab:class_adjustment}
\centering
\begin{tabular}{lcccc}
\toprule
\textbf{Class} &
\multicolumn{1}{c}{\textbf{Auto}} &
\multicolumn{1}{c}{\textbf{Expert}} &
\multicolumn{1}{c}{\textbf{Mean Duration}} &
\multicolumn{1}{c}{\textbf{Mean Duration}} \\
&
\multicolumn{1}{c}{\textbf{Count}} &
\multicolumn{1}{c}{\textbf{Count}} &
\multicolumn{1}{c}{\textbf{Auto (s)}} &
\multicolumn{1}{c}{\textbf{Expert (s)}} \\
\midrule
None  & 929.7 & 825.3 & 0.278 & 0.200 \\
SB    & 400.4 & 314.9 & 0.084 & 0.101 \\
MB    & 42.4  & 61.4  & 0.415 & 0.525 \\
CRS   & 21.3  & 27.2  & 0.531 & 0.625 \\
HS    & 7.3   & 3.0   & 0.740 & 0.307 \\
\bottomrule
\end{tabular}
\end{table}
For the patient dataset, the proposed BS auto-annotation framework correctly captured the relative distribution of bowel sound patterns across all classes. The highest prevalence was consistently observed for the \textit{None} class (Original: 49\%, Predicted: 61\%, Auto: 63.5\%), followed by SB (Original: 40\%, Predicted: 32\%, Auto: 30\%), MB (Original: 6\%, Predicted: 4\%, Auto: 3.5\%), CRS (Original: 4\%, Predicted: 2.5\%, Auto: 2.6\%), and the rarest class, HS (Original: 1\%, Predicted: 0.5\%, Auto: 0.4\%). Importantly, the correct ordering of class prevalence was preserved by the automatic annotation framework.

The auto-annotation algorithm also identified BS events with duration ranges consistent with manual annotations for several patterns. In particular, the shortest-duration patterns—SB (maximum duration $< 0.2\,\mathrm{s}$), CRS (maximum duration $\approx 2\,\mathrm{s}$), and HS (maximum duration $\approx 1\,\mathrm{s}$)—were accurately detected. However, the framework showed a tendency to truncate MB events before their true termination, resulting in systematically shorter detected durations for this class (Original: $3\,\mathrm{s}$; Predicted: $0.9\,\mathrm{s}$; Auto: $1\,\mathrm{s}$).

The most pronounced discrepancy between manual and automatically generated labels was observed for the \textit{None} class for both healthy and patients datasets. This difference arises from the framework’s tendency to explicitly label very short silent intervals between BS events and to segment long silent periods into multiple shorter \textit{None} segments. As a result, the \textit{None} class exhibits an increased distribution and a reduced mean duration in the automatically annotated data compared to the original manual annotations.

The previous evaluation suggests that the BS automatic annotation algorithm can support medical experts by providing an initial characterization of bowel sound signals and enabling a more quantitative assessment. This allows clinicians to quickly measure deviations from normal gastrointestinal motility and to monitor changes over time. Such quantitative information is particularly useful for tracking disease progression, assessing treatment response, and supporting postoperative or critical-care evaluations. Overall, quantitative BS analysis promotes a more standardized and objective interpretation of auscultation findings and facilitates integration into clinical workflows.

\subsubsection*{Use case 2: Human-in-the-loop bowel sound annotation}
The BS auto annotation framework used to annotate the BS of 10 subjects (40 records), and the annotation expert then reviewed the auto-created labels and adjusted them, the required adjustments revealed limited expert intervention. On average, approximately 12\% of detected events were removed or merged, while mean event durations differed by less than 15 ms as it summarized in Tab \ref{tab:class_adjustment}. Class-wise analysis confirmed that the relative distribution of bowel sound patterns was preserved, with corrections primarily affecting event boundaries and the duration of multi-burst events. These results indicate that the proposed auto-annotation framework produces near-expert-quality labels while substantially reducing manual annotation time by 70\%.

The creation of labeled bowel sound datasets is constrained by limited data availability and the high cost of manual annotation, which requires frame-level inspection of long recordings, making it one of the most time-consuming steps in dataset construction. Hence, Automatic pre-annotation combined with expert correction could simplify the annotation of large datasets, offers an efficient strategy for scaling dataset development. 

\section*{Conclusion}
This study presents an automated pipeline for bowel sound segmentation and pattern classification using recordings obtained from a wearable acoustic sensor. The proposed framework combines an energy-based event detection approach with a pretrained deep learning model and applies cohort-specific training to account for differences between healthy subjects and patients.

The results indicate that the proposed algorithm supports multiple use cases, including enabling automatic quantitative analysis of bowel sounds and assisting human annotators during dataset creation. Together, these complementary use cases address both clinical interpretation and the challenge of building high-quality bowel sound datasets, thereby enabling further data-driven analyses that go beyond predefined event characteristics toward a deeper understanding of bowel sounds and their relationship to disease.

\bibliography{sample}

\section*{Acknowledgments}
The authors acknowledge intramural funding from Faculty VI, Carl von Ossiezky Universität Oldenburg (Forschungspool, Potentialbereich mHealth) and from MWK Niedersachsen (via Fraunhofer IDMT, institute part HSA, project Connected Health).

\section*{Data Availability}

Four subjects consented to make their data publicly available. 
The dataset is available at: \url{https://doi.org/10.6084/m9.figshare.28595741.v1}.\\
The full implementation of our approach, along with training scripts, is available in a corresponding code repository \url{https://github.com/AI4HealthUOL/Bowel-sound-auto-annotations}

\end{document}